\documentclass[english,aps,prl,superscriptaddress,twocolumn,address,showacknowledgments,showpacs]{revtex4}
\usepackage[T1]{fontenc}
\usepackage[latin9]{inputenc}
\usepackage{babel}
\usepackage{amsbsy}
\usepackage{amstext}
\usepackage{amssymb}
\usepackage{graphicx}
\usepackage[unicode=true,pdfusetitle,
 bookmarks=true,bookmarksnumbered=false,bookmarksopen=false,
 breaklinks=false,pdfborder={0 0 1},backref=false,colorlinks=false]
 {hyperref}

\makeatletter
\@ifundefined{textcolor}{}
{%
 \definecolor{BLACK}{gray}{0}
 \definecolor{WHITE}{gray}{1}
 \definecolor{RED}{rgb}{1,0,0}
 \definecolor{GREEN}{rgb}{0,1,0}
 \definecolor{BLUE}{rgb}{0,0,1}
 \definecolor{CYAN}{cmyk}{1,0,0,0}
 \definecolor{MAGENTA}{cmyk}{0,1,0,0}
 \definecolor{YELLOW}{cmyk}{0,0,1,0}
}

\makeatother

\begin{document}

\title{Optomechanical Dirac Physics}

\author{M. Schmidt}

\address{University of Erlangen-N\"urnberg, Staudtstr. 7, Institute for Theoretical
Physics, D-91058 Erlangen, Germany}

\author{V. Peano}

\address{University of Erlangen-N\"urnberg, Staudtstr. 7, Institute for Theoretical
Physics, D-91058 Erlangen, Germany}

\author{F. Marquardt}

\address{University of Erlangen-N\"urnberg, Staudtstr. 7, Institute for Theoretical
Physics, D-91058 Erlangen, Germany}

\address{Max Planck Institute for the Science of Light, G\"unther-Scharowsky-Stra\ss e
1/Bau 24, D-91058 Erlangen, Germany}
\begin{abstract}
Recent progress in optomechanical systems may soon allow the realization
of optomechanical arrays, i.e. periodic arrangements of interacting
optical and vibrational modes. We show that photons and phonons on
a honeycomb lattice will produce an optically tunable Dirac-type band
structure. Transport in such a system can exhibit transmission through
an optically created barrier, similar to Klein tunneling, but with
interconversion between light and sound. In addition, edge states
at the sample boundaries are dispersive and enable controlled propagation
of photon-phonon polaritons. 
\end{abstract}


\maketitle
\emph{}

Rapid progress is being made in the field of optomechanics, which
studies the interaction of light with nano-mechanical motion (for
a recent review, see \cite{Aspelmeyer2013RMPArxiv}). Most current
achievements are based on a single vibrational mode coupled to a single
optical mode (i.e. a single ``optomechanical cell''). A logical
next step is to couple many such modes, providing new functionality
and generating new physical phenomena. First steps have been taken
using setups with a few modes (e.g. for synchronization \cite{Zhang2012Sync,Bagheri2013},
wavelength conversion \cite{Hill2012WavelengthConversion,Dong2012},
phonon lasing \cite{Grudinin2010}, or cooling \cite{Bahl2012}).
Going beyond this, we can envisage a periodic arrangement of cells.
In that case we will speak of an ``optomechanical array''. Optomechanical
arrays might be realized on a number of experimental platforms: Microdiscs
\cite{Ding2010,Zhang2012Sync} and microtoroids \cite{Armani2007,Verhagen2012}
could be coupled via evanescent optical fields \cite{Zhang2012Sync}.
Superconducting on-chip microwave cavity arrays (of the type discussed
in \cite{Houck2012}) could be combined with nanobeams \cite{Regal2008}
or membranes \cite{Teufel2011}.\textbf{ }Currently the most promising
platform are optomechanical crystals, i.e. photonic crystals engineered
to contain localized vibrational and optical modes. Single-mode optomechanical
systems based on that concept have been demonstrated experimentally,
with very favorable parameters \cite{Eichenfield2009,Safavi-Naeini2010APL,Gavartin2011PRL_OMC,Chan2011Cooling,SafaviNaeini2014SnowCavity}.
Ab-initio simulations indicate the feasibility of arrays \cite{Safavi-Naeini2011,Heinrich2011CollDyn,Ludwig2013}.
Given these developments it seems that optomechanical arrays are on
the verge of realization.  The existing theoretical work on optomechanical
arrays deals with slow light \cite{Chang2011}, synchronization \cite{Heinrich2011CollDyn,Holmes2012,Ludwig2013},
quantum information processing \cite{Schmidt2012} and quantum many-body
physics \cite{Bhattacharya2008,Tombadin2012,Xuereb2012,Akram2012,Ludwig2013}
and photon transport \cite{Chen2014}. In this letter, we go beyond
these works and illustrate the possibilities offered by engineering
nontrivial optomechanical band structures of photons and phonons in
such arrays. Specifically, we will investigate an array with a honeycomb
geometry. This lattice is the basis for modeling electrons in graphene
\cite{Neto2009}, but it has recently also been studied for photonic
crystals \cite{Peleg2007,Polini2013}, exciton-photon polaritons \cite{Jacqmin2014}
and other systems \cite{Polini2013}. It is the simplest lattice with
a band structure showing singular and robust features called Dirac
cones, mimicking the dispersion of relativistic massless particles.
As we will be interested in the long-wavelength properties of the
structure, on scales much larger than the lattice spacing, we may
call this an ``optomechanical metamaterial''. Tunability would be
the biggest advantage of optomechanical metamaterials, rivaling that
of optical lattices: The band structure is easily tunable by the laser
drive (intensity, frequency, phases). Moreover, it can be observed
by monitoring the emitted light. Using spatial intensity profiles
for driving, one can even engineer arbitrary potentials and hence
local changes in the band structure. We predict that these features
could be used to observe photon-phonon Dirac polaritons, an optomechanical
Klein tunneling effect, and edge state transport. 

\begin{figure}
\includegraphics[width=1\columnwidth]{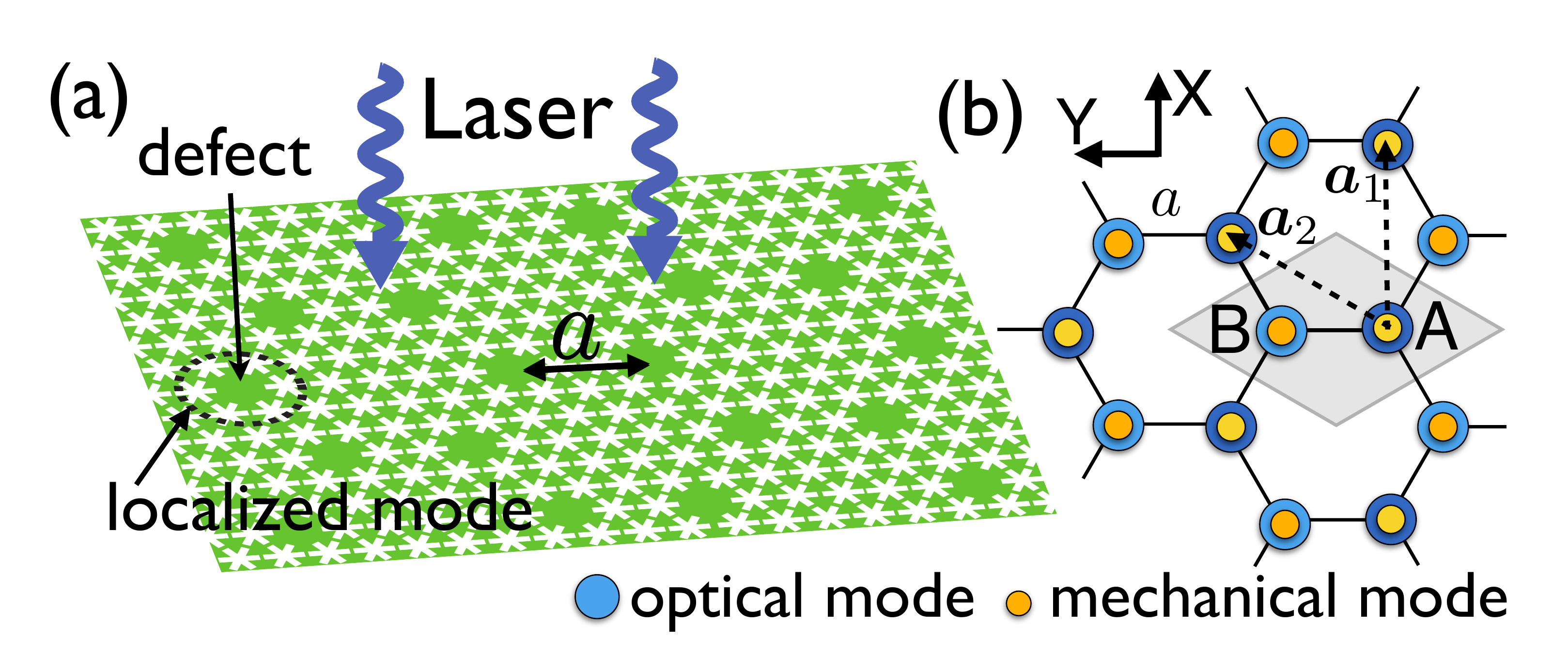}

\caption{\label{fig:Setup}(a) Setup: Thin slabs of free-standing dielectric
(green) with periodically etched holes (white), so-called optomechanical
crystals \cite{Eichenfield2009,Safavi-Naeini2010,Safavi-Naeini2010APL,Gavartin2011PRL_OMC,Chan2011Cooling},
are know to give rise to an optomechanical interaction of localized
optical ($\sim10^{2}\,\text{THz}$) and vibrational modes ($\sim\text{GHz}$)
at engineered defects. The interaction is controlled by a driving
laser. When extended to an array, the modes of nearby defect sites
will be connected via phonon and photon tunneling. (b) We consider
defects arranged in a honeycomb superlattice.}
\end{figure}
\emph{Model -} We consider a 2D honeycomb lattice of identical optomechanical
cells, driven uniformly by a laser (frequency $\omega_{L}$). Each
cell supports a pair of co-localized mechanical (eigenfrequency $\Omega$)
and optical (eigenfrequency $\omega_{{\rm cav}}$) modes interacting
via radiation pressure. This geometry could be implemented based on
optomechanical crystals, see Figure \ref{fig:Setup}, but also in
other physical realizations such as arrays of microdisks, microtoroids,
or superconducting cavities. We adopt the standard approach of linearizing
the dynamics around the steady-state classical solution and performing
the rotating wave approximation, valid for red detuned ($\Delta=\omega_{L}-\omega_{{\rm cav}}<0$)
moderate driving \cite{Aspelmeyer2013RMPArxiv}. In a frame rotating
with the drive, the linearized Hamiltonian reads 
\begin{equation}
\hat{H}/\hbar=\sum_{j}\Omega\hat{b}_{j}^{\dagger}\hat{b}_{j}-\Delta\hat{a}_{j}^{\dagger}\hat{a}_{j}-g_{j}(\hat{b}_{j}^{\dagger}\hat{a}_{j}+\hat{a}_{j}^{\dagger}\hat{b}_{j})+\hat{H}_{{\rm hop}}.\label{eq:LinCell}
\end{equation}
This Hamiltonian describes the non-equilibrium physics of the array
of phonon modes (annihilation operator $\hat{b}_{j}$) and photon
modes ($\hat{a}_{j}$), interacting via the linearized optomechanical
interaction of strength $g_{j}$. The term $\hat{H}_{{\rm hop}}=-\sum(J_{ij}\hat{a}_{i}^{\dagger}\hat{a}_{j}+K_{ij}\hat{b}_{i}^{\dagger}\hat{b}_{j})$
describes the tunneling of photons and phonons between neighboring
sites $i$ and $j$ with amplitudes $J_{ij}$ and $K_{ij}$, respectively
\cite{Heinrich2011CollDyn,Ludwig2013,Safavi-Naeini2011}. Here, $j=[m,n,\sigma]$
is a multi-index, where $m$, $n$ indicate the unit cell, which contains
two optomechanical cells on sublattices A/B (denoted by $\sigma=\pm1$).

The interaction strength is $g_{j}=g_{0}\alpha_{j}$, where $g_{0}$
is the bare optomechanical coupling, i.e. the shift of the local optical
resonance by a mechanical zero-point displacement, and $\alpha_{j}$
is the local complex light field amplitude, proportional to the laser
amplitude \cite{Aspelmeyer2013RMPArxiv}. For completeness, we mention
that the operators $\hat{a}_{j}$ and $\hat{b}_{j}$ in Eq.~(\ref{eq:LinCell})
are assumed shifted, as usual \cite{Aspelmeyer2013RMPArxiv}, by $\alpha_{j}$
and by the radiation-pressure-induced mechanical displacement $\beta_{j}$,
respectively. The detuning $\Delta=\omega_{L}-\omega_{{\rm cav}}$
incorporates a small shift in $\omega_{\text{cav}}$ due to the static
mechanical displacement. 
\begin{figure}
\includegraphics[width=1\columnwidth]{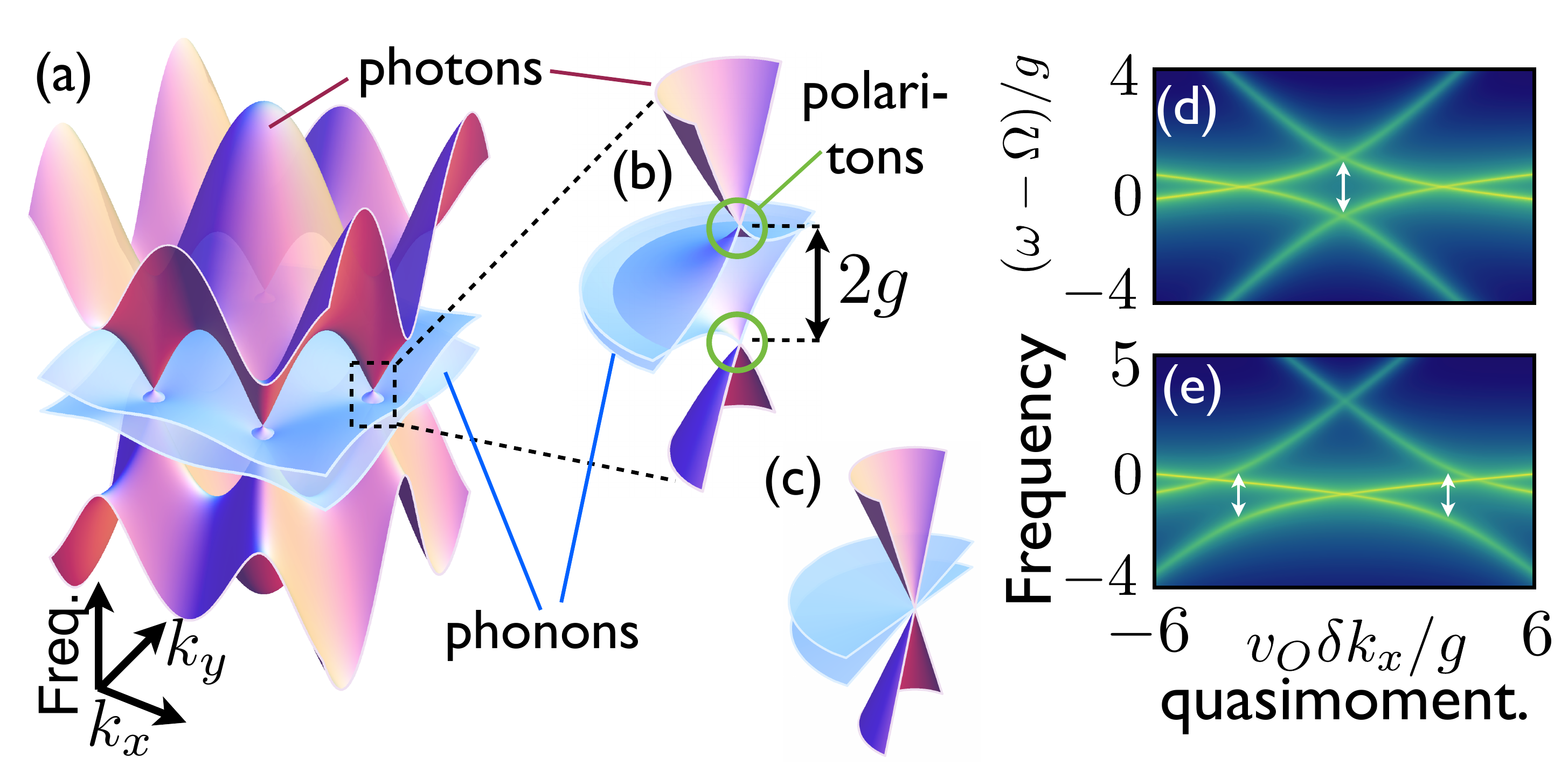}

\caption{\label{fig:OMBandstructure}(a) Band structure of an optomechanical
honeycomb array, featuring fast photons and slow phonons that interact
optomechanically. Detuning the driving laser will shift the photon
band up and down. Here, the photon and phonon Dirac points are chosen
resonant, thus photon-phonon polariton Dirac cones emerge in their
vicinity for $g\neq0$, see the close-up in (b). (c) Without optomechanical
interaction, $g=0$, photon and phonon cones would simply intersect.
(d) A cut through the spectrum $S({\bf k},\omega)$ of the light scattered
by the setup reproduces the band structure, in the presence of dissipation.
(e) Detuned case: Avoided crossing (arrows) between bands with equal
helicity, see main text. {[}Parameters: $v_{M}=v_{O}/10$, $g=J/10$,
$\Delta=-\Omega$ (a-d), $\Delta=-\Omega-3g$ (e), (d,e): $J=\Omega/3$,
$\kappa=J/100$, $\Gamma=\kappa/10$, $\bar{n}=5000${]}}
\end{figure}

The eigenfrequencies of Hamiltonian (\ref{eq:LinCell}) form the optomechanical
band structure, shown in Fig. \ref{fig:OMBandstructure} (a,b) for
realistic parameters and a translationally invariant system ($g_{j}=g$).
It comprises four polariton bands, constructed out of the original
two photon and two phonon bands, giving rise to photon-phonon polariton
Dirac cones. 

A weak additional probe laser can inject excitations at arbitrary
frequency. It can be spatially resolved (via tapered fiber) or momentum-resolved
(extended beam). Even without the probe, the momentum-resolved band
structure is visible in the emitted far-field radiation in the form
of Raman-scattered laser-drive photons, see Fig. \ref{fig:OMBandstructure}
(d,e). We incorporate dissipation and noise via the standard input/output
theory \cite{Aspelmeyer2013RMPArxiv}, taking into account the photon
(phonon) decay rate $\kappa$ ($\Gamma$) and the thermal phonon number
$\bar{n}$, see Supplemental Material. We emphasize that the band
structure (and transport) could be observed in this manner even at
room temperature. 

The emergence of the Dirac cones at the Dirac points ${\bf K}$ and
${\bf K}'$ follows from the symmetries of the honeycomb lattice \cite{Hasan2010RMP}.
Without the drive ($g_{j}=0$), the standard scenario for honeycomb
lattices applies to photons and phonons separately: Excitations can
be on sublattice A or B, corresponding to a binary degree of freedom,
$\sigma_{z}=\sigma=\pm1$. Diagonalizing the Hamiltonian using a plane
wave ansatz, one recovers a $2\times2$ Hamiltonian for every wave
vector ${\bf k}$. Close to a symmetry point, this reduces to the
Dirac Hamiltonian for $2D$ relativistic massless particles. Around
$\mathbf{K}$, it has the form $\hbar v\hat{\boldsymbol{\sigma}}\cdot\delta\mathbf{k}$,
where $\delta\mathbf{k}=\mathbf{k}-\mathbf{K}$ and $\hat{\boldsymbol{\sigma}}$
is the vector of Pauli matrices $\hat{\sigma}_{x,y}$. The photon
velocity at the Dirac point, $v_{O}$, will be generally significantly
larger than the mechanical one, $v_{M}$, see Fig.~\ref{fig:OMBandstructure}(c).
For nearest-neighbor hopping amplitudes $J$ (photons) and $K$ (phonons),
we find $v_{O}=3aJ/2$, $v_{M}=3aK/2$.

We now consider the interacting case ($g\neq0$), turning the Hamiltonian
(\ref{eq:LinCell}) into its first-quantized counterpart in momentum
space and expanding it around a symmetry point. The particle type
can now be encoded in a second binary degree of freedom, $\tau_{z}=\tau=\pm1$
for photons/phonons (with Pauli matrices $\hat{\tau}_{x,z}$). We
find the optomechanical Dirac Hamiltonian: 
\begin{equation}
\hat{H}_{D}/\hbar=\delta\omega\hat{\tau}_{z}/2+(\bar{v}+\delta v\hat{\tau}_{z}/2)\hat{\boldsymbol{\sigma}}\cdot\delta\mathbf{k}-g\hat{\tau}_{x}+\bar{\omega}.\label{eq:OMDirac-1}
\end{equation}
This Hamiltonian describes the mixing of two excitations of very different
physical origin, with properties that are easily tunable. The terms
describe, in this order, an offset between photon and phonon bands,
the Dirac part, and the optomechanical interaction (plus a constant
offset). Here we defined $\bar{v}=(v_{O}+v_{M})/2$, $\delta v=v_{O}-v_{M}$,
$\bar{\omega}=(\Omega-\Delta)/2,$ and $\delta\omega=-\Delta-\Omega$.
The interaction $g$ is tunable in-situ via the drive laser intensity
(in contrast, e.g., to bilayer graphene systems).  Photon-phonon
Dirac polaritons feature a dispersive spectrum
\begin{equation}
\omega_{\tau,\sigma}(\mathbf{k})=\bar{\omega}-\sigma\bar{v}|\delta\mathbf{k}|+\tau\sqrt{g^{2}+(\delta\omega-\sigma\delta v|\delta\mathbf{k}|)^{2}/4},
\end{equation}
 i.~e. the velocity is momentum-dependent and varies on the momentum
scale $g/Ja$, well within the range of validity of Eq.~(\ref{eq:OMDirac-1}),
$\left|\delta\vec{k}\right|\ll a^{-1}$. This effect comes from the
mixing of two Dirac excitations with different velocities.

At the Dirac points, the band structure comprises two pairs of cones
split by $\sqrt{\delta\omega^{2}+4g^{2}}$. Sweeping the laser detuning
$\delta\omega$ from positive to negative values, the upper cones
evolve from purely optical (velocity $v_{O}$), over polaritonic (slope
$\bar{v}=(v_{O}+v_{M})/2$) to purely mechanical (velocity $v_{M}$).
Since the helicity, $\hat{\boldsymbol{\sigma}}\cdot\delta\mathbf{k}/|\delta\mathbf{k}|$,
is conserved, bands of equal helicity feature avoided crossings, while
bands of different helicity cross, see Fig.~\ref{fig:OMBandstructure}(d,e). 

\begin{figure}
\includegraphics[width=1\columnwidth]{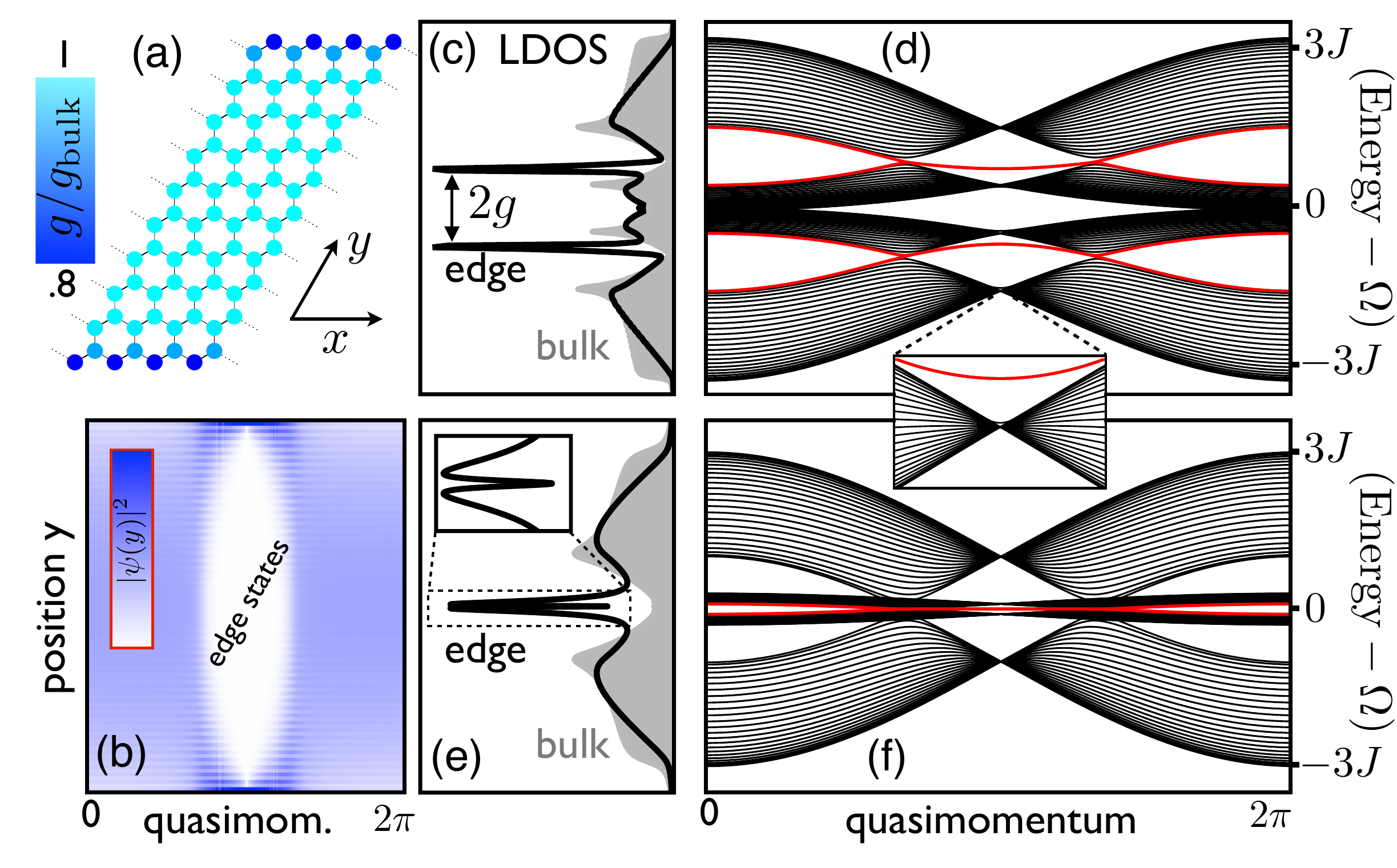}

\caption{\label{fig:EdgeStates}Polariton edge states of a semi-infinite optomechanical
strip (zigzag edge) differ from usual edge states in honeycomb lattices.
(a) Optomechanical interaction strength $g(y)$ of a homogeneously
driven strip. (b) Wavefunction of the upper edge state band. (c) Local
DOS (experimentally accessible via a probe laser) in the bulk (gray)
and at the edge (black) reveals the existence of edge states (here
for $g\gg\kappa$). (d) Corresponding band structure (real part of
eigenfrequencies), indicating the dispersive nature of the edge states
(in red). (e) For $g\ll\kappa$, a sharp dip is observable, due to
optomechanically induced transparency (width $\sim\Gamma$). (f) Band
structure for $g\ll\kappa$. {[}Parameters: $J=\Omega/6$, $K=0.1J$,
$g_{\text{bulk}}=0.007\Omega$ (e,f), $g_{\text{bulk}}=0.15$ (else),
$-\Delta=\Omega$, $\kappa=0.04\Omega$ (e,f), $\kappa=0.01\Omega$
(else), $\Gamma=0.001\Omega$; {]} }
\end{figure}
\emph{Edge states -} The physics of edge states is significantly modified
by inhomogeneous optomechanical couplings that can be tailored via
the laser intensity but also naturally occur in a finite system under
uniform drive. There, the coupling is smaller at the edges than in
the bulk, see Fig. \ref{fig:EdgeStates}(a). In an infinite strip
with zigzag edges this leads to a band of polariton edge modes with
tunable velocity. That is because edge states with momenta closer
to the Dirac points have larger penetration lengths (compare Fig.
\ref{fig:EdgeStates}(b)) and thus explore regions of stronger optomechanical
coupling, making their energy momentum-dependent (Fig.~\ref{fig:EdgeStates}(d)).
In contrast, no transport occurs at the edge of graphene since it
supports a flat band of edge modes \cite{Neto2009}. 

The photonic local density of states (LDOS) is experimentally accessible via
reflection/transmission measurements, e.g. with a tapered fiber probe
brought close to the sample. The LDOS on site $j$, $\rho_{j}(\omega)$,
characterizes the probability to inject a photon with frequency $\omega$.
Figure \ref{fig:EdgeStates}(c) shows the LDOS  for sites in the
bulk (gray) and at the edge (black line). Typical features, like
the vanishing DOS at the Dirac points, are smeared out slightly by
dissipation. The edge states show up as two peaks. For weak coupling
one would naively expect a single edge state peak broadened by dissipation.
However, figure \ref{fig:EdgeStates}(e) shows a peak with a narrow
dip on top. This can be understood as optomechanically induced transparency
\cite{Aspelmeyer2013RMPArxiv}, an interference effect visible for
$\Gamma\ll\kappa$. We note that the gradient in $g$ leads to the
formation of additional bands of edge states, cf. close-up in Fig.
\ref{fig:EdgeStates}(d).

\begin{figure}
\includegraphics[width=1\columnwidth]{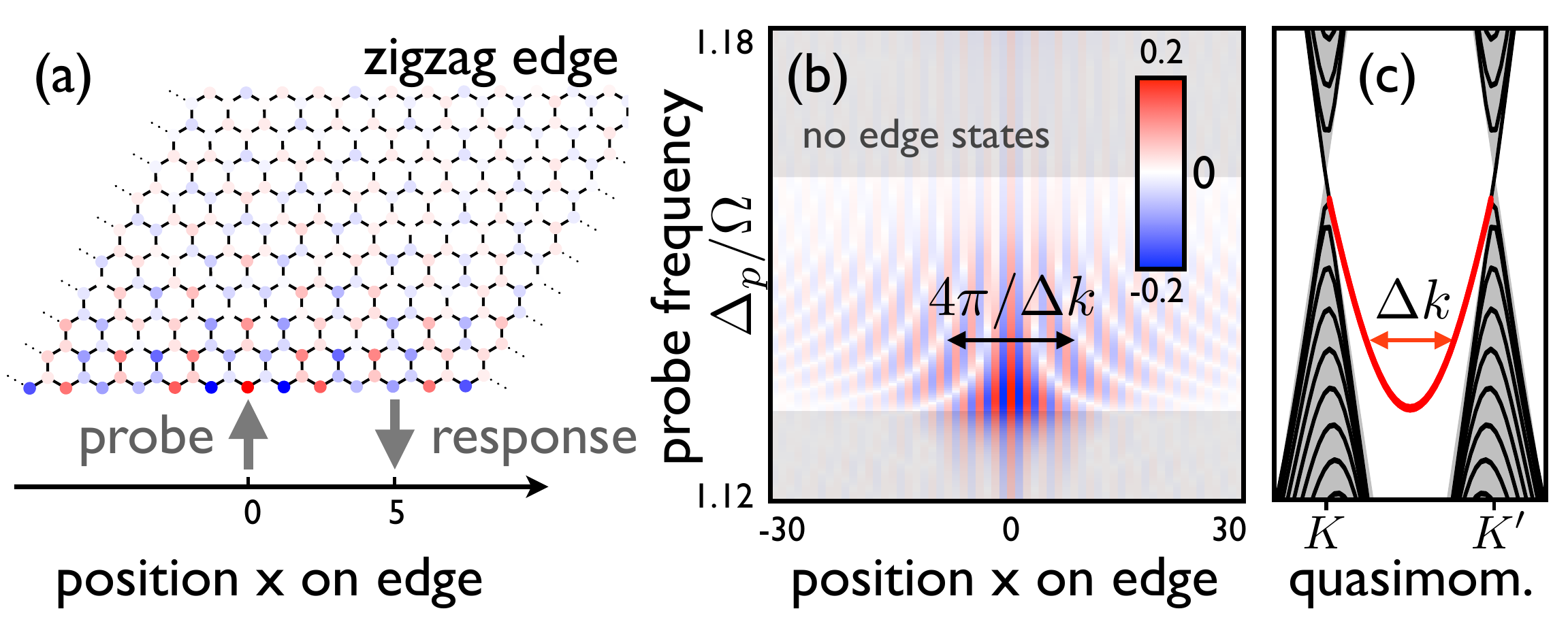}

\caption{\label{fig:Edge-State-Transport}(a) Transport along the edge of a
semi-infinite strip. The optical transmission, $t(\omega,x)$ {[}color
code: $\text{Re}\; t(\omega,x$){]} of a locally injected probe laser.
(b) Real part of the transmission against the probe detuning ($\Delta_{p}=\omega_{\text{probe}}-\omega_{L}$)
and the distance $x$ along the edge. See main text for explanation
of features. The mechanical transmission is proportional in magnitude
to optical one. (c) Close-up of relevant part optomechanical bandstructure.
{[}$g=0.167\Omega$, other parameters as in Fig. \ref{fig:EdgeStates}
(d){]}}
\end{figure}
\emph{Edge state transport }\textendash{} The zigzag edge forms a
polariton waveguide for excitations injected by a local probe at the
edges.  Its group velocity is tunable in-situ via the laser amplitude. Although
the edge states are not protected by a band gap, the transmission
remains mainly along the edge, see Fig. \ref{fig:Edge-State-Transport}(a).
Figure \ref{fig:Edge-State-Transport}(b) depicts the optical transmission
vs. the propagation distance and the probe frequency. For small probe
frequencies there are no edge states, thus the response is local and
weak. Increasing the probe frequency makes edge states resonant, leading
to transmission along the edge. For a given probe frequency, two edge
modes are resonant, with a quasimomentum difference $\Delta k$. This
explains the interference pattern, with transmission minima at $x=\pm n\pi/\Delta k$. The
mechanical transmission mirrors the optical one ($|t_{M}(\omega,x)|\propto|t_{O}(\omega,x)|$)
for strong coupling, and there is no transport for weak coupling (a
flat edge state band). 

\begin{figure}
\includegraphics[width=1\columnwidth]{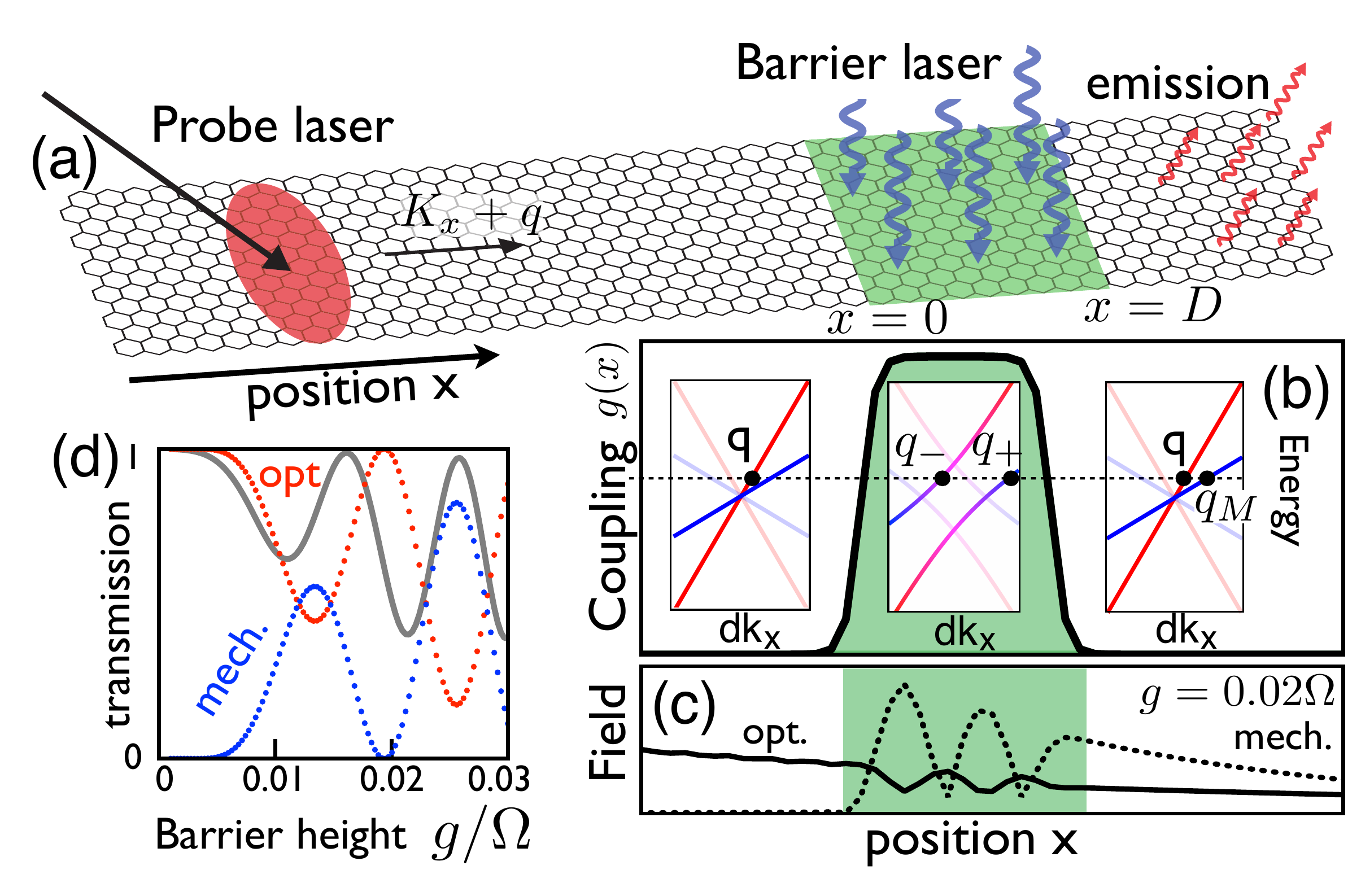}

\caption{\label{fig:Klein-Tunneling}Optomechanical Klein tunneling: (a) A
tilted probe laser injects photons at quasimomentum ${\bf K}+{\bf q}$
that transmit through a barrier (green) as photons and/or phonons
without any backscattering. Emitted light (red arrows) can be detected
experimentally. (b) Position-dependent profile of the optomechanical
coupling $g(x)$, proportional to the light amplitude of the strong
drive laser that creates the barrier. Insets: The local spectrum in
each region, and the allowed quasimomenta at the probe frequency.
(c) Optical and mechanical field ($\left\langle \hat{a}_{j}\right\rangle $
and $\left\langle \hat{b}_{j}\right\rangle $). (d) Optical and mechanical
transmission against barrier height . (gray line: optical transmission
as predicted analytically from the optomechanical Dirac equation)
{[}Parameters:  $\kappa=0.005\Omega$, $\Gamma=0.001\Omega$, $J=\Omega/6$,
$K=J/10${]} }
\end{figure}
\emph{Optomechanical Klein tunneling \textendash{} }The in-situ tunability
of optomechanical metamaterials allows to create arbitrary effective
potential landscapes simply by generating a spatially non-uniform
driving laser profile. This can be nicely illustrated in a setup that
permits the study of Klein tunneling, the unimpeded transmission of
relativistic particles through arbitrary long and high potential barriers.
Electrons in graphene realize a special variant of this \cite{Young2009}. Here,
we show that the backscattering of Dirac polaritons impinging on an
optomechanical barrier is suppressed. Moreover, photons can be converted
into phonons (and vice versa) while being transmitted.

To create a barrier for Dirac photons propagating in the array, we
make use of the distinctive \emph{in-situ} tunability of optomechanical
metamaterials. As shown in Fig.~\ref{fig:Klein-Tunneling}(a), when
a region of width $D$ is illuminated by a strong control laser (of
detuning $\Delta=-\Omega$), a position-dependent optomechanical coupling
$g(x)$ is created. This region represents a barrier for Dirac photons
injected by a probe laser at another spot. We first solve the scattering
problem within the Dirac Hamiltonian (\ref{eq:OMDirac-1}) in the
presence of a barrier with infinitely sharp edges: $g(x)=g$ for $0<x<D$
and $0$ otherwise. We consider a right-moving photon with quasimomentum
perpendicular to the barrier, $|\psi_{{\rm in}}\rangle=e^{iq_{O}x}|\sigma_{x}=1,\tau_{z}=1\rangle$.
Backscattering is forbidden, because the helicity is conserved and
only the right-moving waves {[}bold lines in Fig.~\ref{fig:Klein-Tunneling}(b){]},
have positive helicity $\sigma_{x}=1$. Thus, the wave is entirely
transmitted. Beyond the barrier, it is a superposition of photons
and phonons:
\begin{equation}
\left|\psi_{{\rm out}}\right\rangle =t_{O}e^{iq_{O}x}|1,1\rangle+\sqrt{v_{O}/v_{M}}t_{M}e^{iq_{M}x}|1,-1\rangle,
\end{equation}
where $q_{M}=v_{O}q_{O}/v_{M}$. Note that $|t_{M}|^{2}$ can be
interpreted as the probability that the photon is converted into a
phonon, with $|t_{O}|^{2}=1-|t_{M}|^{2}$ ensuring conservation of
probability. Matching the solutions of the Dirac equation in the different
regions, we find 
\begin{equation}
|t_{M}|^{2}=\sin^{2}[(q_{+}-q_{-})D/2]\big/[1+v_{O}^{3}q_{O}^{2}/(4v_{M}g^{2})],\label{eq:phononconversion}
\end{equation}
where $q_{\pm}$ are the two momenta of the right-moving polaritons
in the interacting region, at the probe frequency. In a more accurate
description,  we compute numerically the stationary light amplitude
$\langle\hat{a}_{j}\rangle$ and the mechanical displacements $\langle\hat{b}_{j}\rangle$
using the full Hamiltonian (\ref{eq:LinCell}) and including also
dissipation, see Supplemental Information. We assume the probe laser
to be injected at a finite distance from the barrier, in a Gaussian
intensity profile, see Fig.~\ref{fig:Klein-Tunneling}(a). The solution,
depicted in Fig.~\ref{fig:Klein-Tunneling}(c), shows all the qualitative
features predicted using the effective relativistic description of
Eq.~(\ref{eq:phononconversion}). Inside the barrier, photons are
converted back and forth into phonons. Phonons reach higher probabilities,
since their speed is smaller ($v_{M}<v_{O}$), and their decay length
is shorter (for realistic parameters $\Gamma/v_{M}>\kappa/v_{O}$).
We deliberately chose a steep barrier (on the scale of the lattice
constant), to illustrate a small Umklapp backscattering to the other
Dirac point (tiny wiggles for $x<0$). The ratio of the phonon current
to the complete current at $x_{0}>D$, $v_{O}\left|a_{0}\right|^{2}/(v_{M}\left|b_{0}\right|^{2}+v_{O}\left|a_{0}\right|^{2})$,
serves as an estimate for the phonon transmission probability $|r|^{2}$.
Figure \ref{fig:Klein-Tunneling}(d) shows the optical and mechanical
transmission against the barrier height, which can be tuned via the
control laser. The fact that the numerical results with dissipation
differ from the theoretical expectation (grey line: $|t_{O}|^{2})$
is mostly due to $v_{M}\ll v_{O}$. Having a large mechanical quasimomentum,
$q_{M}=v_{O}q_{O}/v_{M}\gg q_{O}$, diminishes slightly the quality
of the Dirac approximation. 

\emph{Experimental realizability -} The strong coupling regime, $g>\kappa$,
is routinely reached on several optomechanical platforms, including
optomechanical crystals. It is also crucial to avoid a phonon-lasing
instability, which requires $J\lesssim\Omega/3$ (see Supplemental
Information). In principle, $J$ can be made small by design (e.g.
distance between sites \cite{Heinrich2011CollDyn,Ludwig2013,Safavi-Naeini2011}),
although disorder effects become more pronounced at smaller $J$.
In $2D$, even for frequency fluctuations of the order of $J$, the
Anderson localization length is several hundred sites, safely exceeding
realistic sample sizes. Disorder which is not smooth on the scale
of the lattice constant may still induce Umklapp scattering between
different Dirac points. Numerical simulations indicate that the Klein
tunneling is robust for disorder strengths of $10\%$ of $J$.

\emph{Outlook - }Optomechanical metamaterials will offer a highly
tunable platform for probing Dirac physics using tools distinct from
other systems. Future studies could investigate the rich nonlinear
dynamics expected for blue detuning, which would create novel particle
pair creation instabilities for a bosonic massless Dirac system. Pump-probe
experiments could reveal time-dependent transport processes. Novel
features can also be generated by modifying the laser drive, e.g.
optical phase patterns could produce effective magnetic fields and
topologically nontrivial band structures \cite{Peano2014}, and a
controlled time-evolution of the laser would allow to study adiabatic
changes, sudden quenches and Floquet topological insulators \cite{Oka2009}.

We acknowledge support via an ERC Starting Grant OPTOMECH, via the
DARPA program ORCHID, and via ITN cQOM.

\clearpage

\global\long\def\theequation{S.\arabic{equation}}
 \global\long\def\thefigure{S\arabic{figure}}

\thispagestyle{empty}
\onecolumngrid
\begin{center}
{\fontsize{12}{12}\selectfont \textbf{Supplemental material for the article: ``Optomechanical Dirac Physics''\\[5mm]}}
{\normalsize M Schmidt$^{1}$, V Peano$^{1}$ and F Marquardt$^{1,2}$\\[1mm]}
{\fontsize{9}{9}\selectfont
\textit{	$^1$University of Erlangen-N\"urnberg, Staudtstr. 7, Institute for Theoretical
Physics, D-91058 Erlangen, Germany\\
 			$^2$Max Planck Institute for the Science of Light, G\"unther-Scharowsky-Stra\ss e
1/Bau 24, D-91058 Erlangen, Germany}}
\vspace*{6mm}
\end{center}
\normalsize

\section*{Classical stationary solutions}

In a frame rotating with the driving, the equations of motion for
the classical fields (averaged over classical and quantum fluctuations)
of an optomechanical array read 
\begin{eqnarray}
\dot{\beta}_{\mathbf{j}} & = & (-i\Omega-\Gamma/2)\beta_{j}+ig_{0}|\alpha_{j}|^{2}+i\sum_{\mathbf{l}}K_{\mathbf{jl}}\beta_{l},\nonumber \\
\dot{\alpha}_{\mathbf{j}} & = & (i\Delta^{(0)}-\kappa/2)\alpha_{\mathbf{j}}+i2g_{0}\alpha_{\mathbf{j}}{\rm Re}\beta_{\mathbf{j}}+i\sum_{\mathbf{l}}J_{\mathbf{jl}}\alpha_{\mathbf{l}}+\sqrt{\kappa}\alpha_{L}.\label{eq:stationaryfields-1}
\end{eqnarray}
Here, $\alpha_{L}$ is the laser amplitude and $\Delta^{(0)}=\omega_{L}-\omega_{{\rm cav}}^{(0)}$
is the (bare) detuning. Notice that, in deriving the above equations,
we have just incorporated a general coherent coupling $\hat{H}_{{\rm hop}}$
to the standard equations for single uncoupled optomechanical cells
\cite{Aspelmeyer2013RMPArxiv}. Implicitly, we have assumed that the
dissipation is caused by independent fluctuations on the different
lattice sites. For an infinite array one can readily find a stationary
solution of the classical equations (\ref{eq:stationaryfields-1})
using the mean field ansatz, $\alpha_{\mathbf{j}}=\alpha$ and $\beta_{\mathbf{j}}=\beta$.
The resulting equations have the same form as the equations for single-mode
optomechanics \cite{Meystre1985} 
\begin{equation}
\alpha=\sqrt{\kappa}\alpha_{L}/[\Delta^{(0)}+2g_{0}\beta-\nu_{O}+i\kappa/2],\qquad\beta=g_{0}|\alpha|^{2}/(\Omega+\nu_{M}).\label{eq:stationaryoscillations}
\end{equation}
As in the standard case, the radiation pressure induced mechanical
displacement $\beta$ translates into a shift of the optical mode
eigenfrequencies, $-2g_{0}\beta$. In the main text, we incorporate
this shift in the effective detuning $\Delta=\Delta^{(0)}+2g_{0}\beta$.
An additional shift of the mechanical and optical eigenfrequencies
is induced by the coupling to the neighboring sites, $\nu_{O}=-\sum_{\mathbf{l}}J_{\mathbf{jl}}$
and $\nu_{M}=-\sum_{\mathbf{l}}K_{\mathbf{jl}}$ (for nearest neighbor
hopping $\nu_{O}=3J$ and $\nu_{M}=3K$). For a finite array the stationary
fields $\alpha_{\mathbf{j}}$ and $\beta_{\mathbf{j}}$ are not independent
of $\mathbf{j}$ . In this case, we solve the classical equations
(\ref{eq:stationaryfields-1}) numerically.

\section*{Linearized Langevin equations}

In our work, we apply the standard approach of linearizing the dynamics
around the classical solutions \cite{WallsMilburn_QuantumOptics},
the linearized Langevin equations read
\begin{eqnarray}
\dot{\hat{b}}_{j} & = & i\hbar^{-1}[\hat{H'},\hat{b}_{\mathbf{j}}]-\Gamma\hat{b}_{\mathbf{j}}/2+\sqrt{\Gamma}\hat{b}_{\mathbf{j}}^{({\rm in)}}=(-i\Omega-\Gamma/2)\hat{b}_{\mathbf{j}}+ig_{\mathbf{j}}\hat{a}_{\mathbf{j}}+ig_{\mathbf{j}}\hat{a}_{\mathbf{j}}^{\dagger}+i\sum_{\mathbf{l}}K_{\mathbf{jl}}\hat{b}_{\mathbf{l}}+\sqrt{\gamma}\hat{b}_{\mathbf{j}}^{({\rm in)}},\nonumber \\
\dot{\hat{a}}_{j} & = & i\hbar^{-1}[\hat{H}',\hat{a}_{\mathbf{j}}]-\kappa\hat{a}_{\mathbf{j}}/2+\sqrt{\kappa}\hat{a}_{\mathbf{j}}^{({\rm in)}}=(i\Delta_{\mathbf{j}}-\kappa/2)\hat{a}_{\mathbf{j}}+ig_{\mathbf{j}}(\hat{b}_{\mathbf{j}}+\hat{b}_{\mathbf{j}}^{\dagger})+i\sum_{\mathbf{l}}J_{\mathbf{jl}}\hat{a}_{\mathbf{l}}+\sqrt{\kappa}\hat{a}_{\mathbf{j}}^{({\rm in})}\label{eq:langevin}
\end{eqnarray}
with the noise correlators
\begin{eqnarray}
\langle\hat{a}_{\mathbf{j}}^{(in)}(t)\hat{a}_{\mathbf{l}}^{(in)\dagger}(0)\rangle & = & \kappa\delta_{\mathbf{jl}}\delta(t),\qquad\langle\hat{a}_{\mathbf{j}}^{(in)\dagger}(t)\hat{a}_{\mathbf{l}}^{(in)}(0)\rangle=0,\nonumber \\
\langle\hat{b}_{\mathbf{j}}^{(in)}(t)\hat{b}_{\mathbf{l}}^{(in)\dagger}(0)\rangle & = & \Gamma(\bar{n}+1)\delta_{\mathbf{jl}}\delta(t),\qquad\langle\hat{b}_{\mathbf{j}}^{(in)\dagger}(t)\hat{b}_{\mathbf{l}}^{(in)}(0)\rangle=\Gamma\bar{n}\delta_{\mathbf{jl}}\delta(t).\label{eq:noisecorrelatorsdef}
\end{eqnarray}
 The output fields are related to the fields in the array and the
input fields by the input output relations \cite{WallsMilburn_QuantumOptics},
\begin{equation}
\hat{a}_{\mathbf{j}}^{({\rm out})}=\hat{a}_{\mathbf{j}}^{({\rm in})}-\sqrt{\kappa}\hat{a}_{\mathbf{j}},\quad\hat{b}_{\mathbf{j}}^{({\rm out})}=\hat{b}_{\mathbf{j}}^{({\rm in})}-\sqrt{\Gamma}\hat{b}_{\mathbf{j}}.\label{eq:inputoutput}
\end{equation}

Notice that $\hat{H}'=\hat{H}+\hat{H}_{{\rm st}}$ contains also counter
rotating terms, $\hat{H}_{st}=\sum_{\mathbf{j}}g_{\mathbf{j}}\left(\hat{a}_{\mathbf{j}}^{\dagger}\hat{b}_{\mathbf{j}}^{\dagger}+\hat{a}_{\mathbf{j}}\hat{b}_{\mathbf{j}}\right).$
These terms have been omitted in Eq. (1). This is the standard rotating
wave approximation which applies to any side band resolved optomechanical
system driven by a red detuned laser with a moderate intensity, $\Omega\gg\kappa$
and $g^{2}\lesssim\kappa\Omega$. In an optomechanical array, the
laser should be red detuned compared to the lowest frequency optical
eigenmode. Thus, in the regime when Dirac photons and Dirac phonons
are resonantly coupled ($-\Delta\approx\Omega$), we find the additional
constraint $J<\Omega/3$ .

\subsection*{Photon emission spectrum}

In Fig. 2(d,e), we plot the power spectrum $S(\mathbf{k},\omega)$
of the photons emitted by the array (periodic boundary conditions
have been assumed), 
\begin{equation}
S(\mathbf{k},\omega)\equiv\sum_{\sigma}\int dt\exp[i\omega t]\langle\hat{a}_{\mathbf{k}\sigma}^{\dagger}(t)\hat{a}_{\mathbf{k}\sigma}\rangle.\label{eq:powerspectrumdef}
\end{equation}
Here, $\hat{a}_{\mathbf{\mathbf{k}\sigma}}$ are the annihilation
operators of the photonic Bloch modes, $\hat{a}_{\mathbf{\mathbf{j}}}=({\cal N})^{-1/2}\sum_{\mathbf{j}}e^{i\mathbf{k}\cdot\mathbf{r}_{j}}\hat{a}_{\mathbf{k}\sigma}$
{[}$\mathbf{r_{j}}$ is the position counted off from a site on sublattice
$A$ and ${\cal N}$ is the number of unit cells{]}. In a large array
(where finite size effects are smeared out by dissipation), $S(\mathbf{k},\omega)$
is proportional to the angle-resolved radiation emitted by the array
at frequency $\omega_{L}-\omega$.

For periodic boundary conditions and nearest neighbor hopping, the
Langevin equations (\ref{eq:langevin}) can be solved analytically
(including also the counter rotating terms). By plugging the corresponding
solutions into the definition (\ref{eq:noisecorrelatorsdef}) and
taking into account the correlators Eqs. (\ref{eq:noisecorrelatorsdef}),
we find
\begin{equation}
S(\mathbf{k},\omega)=\sum_{\sigma}\frac{4\kappa g^{4}\Omega^{2}+\Gamma\sigma_{{\rm M}}(\omega,\Delta(\mathbf{k},\sigma),\Omega(\mathbf{k},\sigma))}{|{\cal N}(\omega,\Delta(\mathbf{k},\sigma),\Omega(\mathbf{k},\sigma))|^{2}}\label{eq:powerspectrum}
\end{equation}
in terms of the analytical functions
\begin{eqnarray*}
 &  & \sigma_{{\rm M}}(\omega,\Delta,\Omega)=g^{2}|\chi_{_{O}}(\omega,\Delta)|^{-2}\left[(\bar{n}+1)|\chi_{_{M}}(-\omega,\Omega)|^{-2}+\bar{n}|\chi_{_{M}}(\omega,\Omega)|^{-2}\right]\\
 &  & {\cal N}(\omega,\Delta,\Omega)=[\chi_{_{O}}(\omega,\Delta)\chi_{_{M}}(\omega,\Omega)\chi_{_{O}}^{*}(-\omega,\Delta)\chi_{_{M}}^{*}(-\omega,\Omega)]^{-1}+4g^{2}\Delta\Omega.
\end{eqnarray*}
Here, we have introduced the free susceptibilities $\chi_{_{O}}(\omega,\Delta)=[\kappa/2-i(\omega+\Delta)]^{-1}$
and $\chi_{_{M}}(\omega,\Omega)=[\Gamma/2-i(\omega-\Omega)]^{-1}$.
Moreover, $-\Delta(\mathbf{k},\sigma)$ and $\Omega(\mathbf{k},\sigma)$
are the spectra of tight-binding photons and phonons on the honeycomb
lattice (the photon spectrum is defined in the rotating frame), respectively.
They are given by $\Delta(\mathbf{k},\sigma)=\Delta+Jf(\mathbf{k},\sigma)$
and $\Omega(\mathbf{k},\sigma)=\Omega-Kf(\mathbf{k},\sigma)$ where
$f(\mathbf{k},\sigma)=\pm|1+e^{i\mathbf{k}\cdot\mathbf{a}_{1}}+e^{i\mathbf{k}\cdot\mathbf{a}_{2}}|$.

\subsection*{Local Density of states and transmission amplitudes}

In Fig. 3 and 4 of the main text, we plot the local photonic densities
of states (LDOS) on site $\mathbf{j}$, $\rho(\omega,\mathbf{j})$
and the transmission amplitude $t_{O}(\omega,\mathbf{l},\mathbf{j})$
relating the emission in the output field at site $\mathbf{l}$ to
an input probe field at sites $\mathbf{j}$ with frequency $\omega$,
$\left\langle \hat{a}_{\mathbf{l}}^{({\rm out})}(t)\right\rangle =t_{O}(\omega,\mathbf{l},\mathbf{j})\left\langle \hat{a}_{\mathbf{j}}^{({\rm in})}(t)\right\rangle $
where $\left\langle \hat{a}_{\mathbf{j}}^{({\rm in})}(t)\right\rangle =fe^{-i\omega t}$.
These two quantities are directly related to the photonic retarded
Green's function
\[
\tilde{G}_{OO}(\omega,\mathbf{j},\mathbf{l})=-i\int_{-\infty}^{\infty}dte^{i\omega t}\Theta(t)\langle[\hat{a}_{\mathbf{j}}(t),\hat{a}_{\mathbf{l}}^{\dagger}(0)]\rangle.
\]
 In fact, the density of state is defined as
\begin{equation}
\rho(\omega,\mathbf{j})=-2{\rm Im}\tilde{G}_{O}(\omega,\mathbf{j},\mathbf{j})\label{eq:ldos}
\end{equation}
where $\tilde{G}_{OO}(\omega,\mathbf{j},\mathbf{l})=-i\int_{-\infty}^{\infty}dte^{i\omega t}\Theta(t)\langle[\hat{a}_{\mathbf{j}}(t),\hat{a}_{\mathbf{l}}^{\dagger}(0)]\rangle$.
Moreover, using Kubo formula and the input output relation Eq. (\ref{eq:inputoutput}),
we find the photon transmission amplitude to be 
\begin{equation}
t_{O}(\omega,\mathbf{l},\mathbf{j})=\delta_{\mathbf{lj}}-i\kappa\tilde{G}_{OO}(\omega,\mathbf{l},\mathbf{j}).\label{eq:transmission}
\end{equation}

For an infinite strip of width $M$ unit cells, it is most convenient
to introduce the partial Fourier transform of $\tilde{G}(\omega,\mathbf{j},\mathbf{l})$,
\begin{equation}
\tilde{G}_{OO}(\omega,\mathbf{j},\mathbf{l})=N^{-1}\sum_{k}e^{i(n_{j}-n_{l})k_{x}}\tilde{G}_{OO}(\omega,k_{x};m_{j},\sigma_{J};m_{l},\sigma_{l}).\label{eq:Greenmomentum}
\end{equation}
Here, $k_{x}$ is the momentum in the translationally invariant direction
($x$-axis). Formally, we have introduced a finite length of $N$
cells and periodic boundary conditions. However, the spurious finite
size effects induced by this assumption are smeared out by dissipation
for an appropriately large $N$. After taking the partial Fourier
transform of the classical displaced fields $\langle\hat{a}_{\mathbf{j}}\rangle$
and $\langle\hat{b}_{\mathbf{j}}\rangle$, we organize their Fourier
components $\alpha_{k_{x}m\sigma}$, $\beta_{k_{x}m\sigma}$ in a
$2M$-dimensional vector $\mathbf{c}_{k}$ with equation of motion
in the form $i\langle\dot{\hat{\mathbf{c}}}_{k}\rangle=A_{k}\langle\hat{\mathbf{c}}_{k}\rangle$
(when no probe laser is present). The $2M\times2M$ matrix $A_{k}$
is obtained from the Langevin equations (\ref{eq:langevin}) by neglecting
the counter rotating terms. Thus, the Green's function $\tilde{G}_{OO}(\omega,k_{x};m_{j},\sigma_{J};m_{l},\sigma_{l})$
is the block of the matrix $\tilde{G}(\omega,k)=(\omega-A_{k})^{-1}$
which acts on the optical subspace of $\hat{\mathbf{c}}_{k}$. The
LDOS and transmission amplitudes $t(\omega,\mathbf{i},\mathbf{j})$
are then readily calculated from Eqs. (\ref{eq:ldos}-\ref{eq:Greenmomentum})

\section*{Details of the numerical calculation of the Klein tunneling of photons
and phonons}

In Fig. 5, we consider an infinite strip with armchair edges and a
width of $N=500$ unit cells (in the $x$-direction). Notice that
the unit cell of an armchair strip is formed by four sites. Thus,
the photon and phonon dynamics is described by the Langevin equations
(\ref{eq:langevin}) with the multi-index $\mathbf{j}=[m_{x},m_{y},s]$,
where $m_{x}=0,\ldots,N$, $m_{y}\in Z$, and $s=1,2,3,4$. The optomechanical
barrier created by the strong control laser $ $is translationally
invariant in the $y$-direction, $g(m_{x})=g\left[e^{\beta(m_{x}-m_{R})}+1\right]^{-1}\left[e^{\beta(m_{L}-m_{x})}+1\right]^{-1}$
with $\beta=2$, $m_{L}=200$, and $m_{R}=213$ . The probe laser
has a gaussian intensity profile in the $x$-direction with average
inplane momentum close to the $\mathbf{K}$ symmetry point, $\hat{a}_{\mathbf{j}}^{({\rm in)}}=\exp[-i\Delta_{p}t-(m_{x}-m_{0})^{2}/\delta m^{2}+i\mathbf{r_{j}\cdot}\mathbf{\bar{k}}].$
We choose $\bar{\mathbf{k}}-\mathbf{K}=(0.029/a,0)$, $\Delta_{p}=\Omega+v_{O}|\bar{\mathbf{k}}-\mathbf{K}|$,
$m_{0}=90$, and $\delta m=30$. The other parameters are given in
the main text. The stationary Langevin equations have been solved
by computing numerically the Green's functions for $k_{y}=0$.

\end{document}